\documentclass[a4paper,fleqn,usenatbib,useAMS]{mnras}
\usepackage{pdfpages}
\usepackage{graphicx}	\usepackage{amsmath}	\usepackage{amssymb}	\usepackage{multicol}        \usepackage{bm}		\usepackage{pdflscape}	\usepackage{arydshln}

\usepackage[T1]{fontenc}
\usepackage{ae,aecompl}

\usepackage{txfonts}

\title[Gravitational energy spectrum of molecular cloud]{Constructing
multiscale gravitational energy spectra from  molecular cloud surface density PDF -- Interplay between turbulence and
gravity}
\author[Guang-Xing Li,  Andreas Burkert]{Guang-Xing Li$^{1}$\thanks{Contact e-mail:
\href{mailtogx}{gxli@usm.lmu.de}} ,  Andreas
Burkert$^{1, 2}$
\\
$^{1}$University Observatory Munich, Scheinerstrasse 1, D-81679 M\"unchen,
Germany\\
$^{2}$ Max-Planck-Fellow, Max-Planck-Institute for Extraterrestrial
Physics, Giessenbachstrasse 1, 85758 Garching, Germany
}

\date{\today}

\pubyear{2016}

\begin{document}
\label{firstpage}
\pagerange{\pageref{firstpage}--\pageref{lastpage}}
\maketitle

\begin{abstract}
Gravity is believed to be important on multiple physical scales
in molecular clouds. However, quantitative constraints on gravity are still lacking.
We derive
an analytical formula which provides estimates on multiscale gravitational
energy distribution  using the observed surface density PDF.
Our analytical formalism also enables one to convert the observed column density
PDF into an estimated volume density PDF,  and to obtain average radial density
profile $\rho(r)$.
For a region with $N_{\rm col} \sim N^{-\gamma_{\rm N}}$, the gravitational
energy spectra is $E_{\rm p}(k)\sim k^{-4(1 - 1/\gamma_{\rm
N})}$. We apply the formula to observations of
molecular clouds, and find that a scaling index of $-2$ of the surface
density PDF implies that $\rho \sim r^{-2}$ and $E_{\rm p}(k)  \sim k^{-2}$.
{ The results are valid from the cloud scale (a few parsec) to around
$\sim 0.1 \;\rm pc$. }Because of the resemblance the scaling index of the
gravitational energy spectrum and the that of the kinetic energy power spectrum of the Burgers
turbulence (where $E\sim k^{-2}$), our result indicates that gravity can act effectively
against turbulence over a multitude of physical scales.
This is the critical scaling index which divides molecular clouds into two
categories:
 clouds like
Orion and Ophiuchus have shallower power laws, and the amount of gravitational
energy is too large for turbulence to be effective inside the cloud. Because
gravity dominates, we call this type of cloud \textit{g-type} clouds.
On the other hand, clouds like the California molecular cloud and the Pipe nebula have
steeper power laws, and turbulence can overcome gravity if it can cascade
effectively from the large scale.  We call this type of cloud \textit{t-type}
clouds.
The analytical formula can be used to determine if gravity is dominating cloud
evolution when the column density probability distribution function (PDF) can be
reliably determined.
  \end{abstract}


\begingroup
\let\clearpage\relax
\endgroup
\newpage

\section{Introduction}
The interplay between turbulence and gravity plays determining roles in
the dynamics of astrophysical fluids, and in particular molecular clouds
\citep[][and references therein]{2014prpl.conf....3D}.
Turbulence is a self-similar process.
It is believed that it can effectively transport energy from the large scale to
 smaller scales, and provide necessary support to stop matter from
collapsing too rapidly \citep{2004RvMP...76..125M}. The effect of turbulence in star formation
is subject to intensive investigations during the past decade (see e.g.
\citet{2007prpl.conf...63B} and references therein).

Gravity, being the only long-range and attractive force, determines the dynamics
of the majority of  astrophysical systems. Molecular clouds exhibit
structures on a multitude of physical scales -- from at least $10^2$ parsec down
to sub-parsec scales \citep{2000prpl.conf...97W}.
Because gravity is scale-free, it can be important on all these physical scales.
Previously, the importance of gravity has been quantified on various scales
using the virial parameter \citep{1992ApJ...395..140B}. The importance of
gravity on multiple physical scales has also been observationally
demonstrated \citep{2009Natur.457...63G,2015A&A...578A..97L}. One limitation of
the virial parameter is that one needs to specify a scale on
which it can be evaluated. Thus it is difficult to obtain a
multiscale picture of gravity based on the virial parameter alone. Besides, the
virial parameter is not additive. Thus it is not useful for evaluating the
importance of gravity on bulk molecular gas.

A better representation of the importance of gravity is the gravitational energy.
Compared to the virial parameter, energy is additive. One can separate the
total gravitational energy of a molecular cloud into contributions from
different parts and from different physical scales. This provides a
multiscale picture of gravity in molecular clouds -- an important piece of information that is still missing.
This task is now feasible as observations can reliably trace the gas from $10^2$
parsec down to sub-parsec scales
e.g.
\citet[][]{2013ApJ...766L..17S,2012A&A...540L..11S,2009A&A...508L..35K,2015A&A...576L...1L}.

The structure of a molecular cloud can be interpreted by the surface density
PDFs (PDF is the probability distribution function). It is a measure of the
distribution of the observed surface density structure of a molecular cloud.
Observationally, the PDFs are found to exhibit power-law forms
$P_{\rm N_{\rm col}}\sim N_{\rm col}^{-\gamma_{\rm N}}$
\citep[][]{2015A&A...576L...1L,2013ApJ...766L..17S,2012A&A...540L..11S,2009A&A...508L..35K,2015A&A...576L...1L,2014Sci...344..183K},
at least for the parts with high surface densities \citep{2009A&A...508L..35K}. This is usually interpreted
as the system being strongly self-gravitating, perhaps also influenced by
rotation and magnetic field
\citep{2011ApJ...727L..20K,2014ApJ...781...91G,2009A&A...508L..35K}.
{ We note, however, that power-law PDFs has been noticed in other
some earlier simulations. See e.g.
\citet{1998ApJ...504..835S,2008PhST..132a4025F,2008MNRAS.390..769V,2000ApJ...535..869K,2011ApJ...731...59C}.}
One importance piece of information that one can extract from an observed
surface density PDF is to constrain how the gravitational energy of a molecular cloud is distributed across different physical scales. This is the major focus of this work
\footnote{ The reader might be interested in other methods that
quantifies the cloud structures, e.g. the correlation function
\citep{2013ApJ...763...51F,2015ApJ...808...48B,2012ApJ...750...13C},
potential-based G-virial method \citep{2015A&A...578A..97L}, and Dendrogram
method \citep{2009Natur.457...63G,2008ApJ...679.1338R}. A thorough comparison
can be found in an accompanying paper \citep{2016arXiv160304342L}.}.

We present observational constraints on the multiscale
importance of gravity by inferring it from the observed surface density PDF of
a molecular cloud. The formalism is based on a simple view, that the
high-density parts of the gas tend to be surrounded by gas of
relatively lower densities.
This insight enables us to construct a nested \emph{shell model} for the dense parts of
 molecular clouds. The model characterises the structures  seen in
 observations and yet at the same time enables us to evaluate the contribution
 to the gravitational energy from various physical scales with an analytical
 approach.
The paper is organized as follows: In Section \ref{sec:model} we describe our
model, and derive an analytical formula to compute the multiscale gravitational energy
distribution (called \emph{gravitational energy spectrum} in this work) from the
observed surface density PDF. The formulas to convert the observed column
density PDF into volume density PDF, averaged radial profile and gravitational
energy spectra are summarized in Sec. \ref{sec:formulas}. Then these formulas
are applied to observations to provide constraints (Sec. \ref{sec:obs}).
In Sec.
\ref{sec:conclu} we conclude.

\section{multiscale gravitational energy}\label{sec:model}

\subsection{Observational picture}
A surface density PDF is a statistical probability distribution of the observed
surface densities of a molecular cloud.  At high surface densities, molecular
clouds exhibit power-law PDFs.
It has been demonstrated that one can ``unfold'' the
observed surface density distribution into the intrinsic density PDF
($\rho$-PDF, \citep{2014Sci...344..183K}), either with
a volume density modelling technique \citep{2014Sci...344..183K} or with an
analytical formula
\citep{2014ApJ...781...91G,2011ApJ...727L..20K,2013ApJ...763...51F,2010MNRAS.405L..56B}.

Suppose that above a critical surface density of $N_{\rm col,\ min}$, a
molecular cloud has an observed surface density distribution of  \footnote{Here,
the PDF are normalized with the observed area, i.e., $P(N_{\rm col})$ has a
dimension of $A=L^2$ where $L$ is the size of the region. Similarly, the
rho-PDF $P(\rho)$ have a dimension of volume $V = L^3$.   }
\begin{equation}
P(N_{\rm col}) = P_{N_0} \Big( \frac{N_{\rm col}}{N_{0}} \Big)^{- \gamma_{\rm
N}}\;,
\end{equation}
where $N_{\rm col}$ is the observed surface density and $\gamma_{\rm N} \approx
2$. {  This is a fiducial value, and observationally,
different clouds have very different slopes. For star-forming cloud,
$\gamma_{\rm N}$ can reach 1.5; for non-star-forming clouds,  $\gamma_{\rm
N}\approx 4$. The range of scaling exponents has been seen in both
extinction-based observations e.g. \citep{2009A&A...508L..35K} and in
emission-based measurements e.g.
\citep{2012A&A...540L..11S,2014A&A...566A..45L}.
 See also Table
\ref{tbl:properties}}.
The $\rho$-PDF can be estimated as \citep{2014ApJ...781...91G} \begin{equation}\nonumber P(\rho) = P_{\rho_0} \Big(\frac{\rho}{\rho_0}\Big)^{- \gamma_{\rho}}\;,
\end{equation}
where
\begin{equation}\label{eq:ncoltorho}
\gamma_{\rho} = \frac{3 \gamma_{\rm N}}{\gamma_{\rm N} + 2}\;.
\end{equation}
where $\rho_0 \approx N_0 / L$ and $L$ is the size of the region.
{  The normalization depends on $L$, which one can measure form
the images, and the normalization is accurate only in order-of-magnitude sense.
} As a crude estimate, $L \approx \sqrt{P_{N_0} }{\rm d}x$ where ${\rm d} x $ is the pixel size.
On the 2D
plane, the region is composed of $M \approx N\times N = (L/{\rm d}x)^2$
pixels which implies $P_{N_0} \approx L^2$, and in 3D, the region is
composed of $M' \approx N\times N \times N= (L/{\rm d}x)^3$ vorxels, which
implies $P_{\rho_0} \approx L^3$. Here, $P_{N_0}$  and $P_{\rho_0}$ have
dimensions of area and volume, respectively.
 As has been demonstrated in
\citet{2014ApJ...781...91G,2013MNRAS.436.3247K,2014A&A...565A..24F}, this
empirical relation is reasonably accurate when applied to numerical simulations
where a multitude of structures are present.

\subsection{The \emph{shell model}}
\begin{figure*}
\includegraphics[width = 0.9 \textwidth]{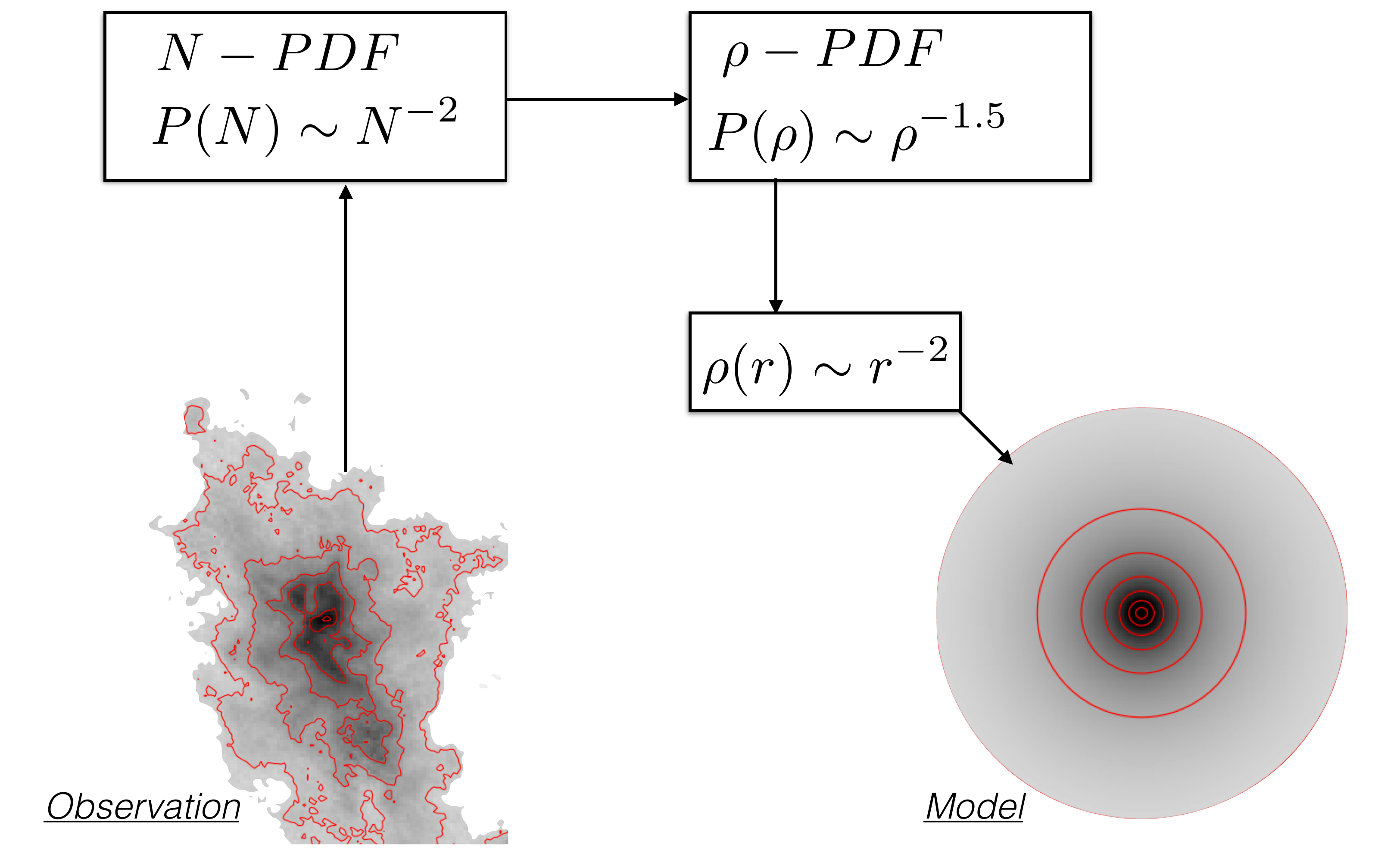}
\caption{A illustration of the model. For a given star-forming region (e.g.
the NGC1333 star forming region as has been shown on the left), we can
constraint is surface density PDF (N-PDF) observationally. Then we can construct
a volume density PDF ($\rho$-PDF) based on the surface density PDF. Finally, we
 construct a \emph{shell model} where the density distribution can be described
 as the effective radial profile $\rho(r) \sim r^{-2}$ where $r$ is the radius.
 This is an approximation to the real density structure of the region. Turbulence can be driven from the outer shell
and would cascades inwards. The image of NGC1333 (on the left) is produced from
the {  velocity-integrated} $^{13}$CO(1-0) data from the COMPLETE
survey \citep{2006AJ....131.2921R}. { The image on the right is produced
with the an analytical formula, which can be found in the clump model section of
\citet{2016arXiv160305720L}.} The conversion between density PDF and effective
radial density profiles can be found in Sec.
\ref{sec:formulas}\label{fig:model}.}
\end{figure*}
We assume a multiscale spherical symmetric  nested model for molecular clouds
where gas with high densities stays inside regions of lower densities.  We
consider a simplified model where the entire star-forming region can be approximated as one such structure. We
call this a ``\emph{shell model}''. The basic idea of this simplification is
sketched in Fig.
\ref{fig:model}.

We assume that above a critical density $\rho_{\rm crit}$, a star-forming region
obeys
\begin{equation}\label{eq:prho}
P(\rho)  = P_{\rho_0} \times \Big(\frac{\rho}{\rho_0}\Big)^{- \gamma_{\rho}}\;,
\end{equation}
where a fiducial value of  $\gamma_{\rho}$ woudl be $1.5$. $P_{\rho_0} \approx
L^3$ where $L$ is the size of the region, and  $\rho_0 \approx N_0 / L$.
Here, the probability is measured in terms of surface area, and $P_{\rho_0}$
has a unit of $L^3$ where $L$ is the size. The amount of mass ${\rm d}
M(\rho)$ contained in a shell of mass with $(\rho, \rho + {\rm d} \rho) $ is
\begin{equation}\label{eq:d:mass}
{\rm d} M(\rho) = P(\rho) {\rm d}\rho = P_{\rho_0} \times \Big(\frac{\rho}{\rho_0}\Big)^{- \gamma_{\rho}} {\rm
d}\rho
\end{equation}
and the \emph{enclosed mass} inside such a shell is
\begin{eqnarray}\label{eq:mass}
 M_{\rm enc}(\rho) &=& \int_{\rho}^{\rho_{\rm max}} P(\rho') {\rm d}\rho' \\
 \nonumber &=&
 P_{\rho_0} \rho_0 \frac{(\rho' / \rho_0)^{1 - \gamma_{\rho} }}{1 -
 \gamma_{\rho}}\Big{|}^{\rho_{\rm max}}_{\rho} = \frac{P_{\rho_0}\rho_0 }
 {\gamma_{\rho} - 1}\Big{(}\frac{\rho}{\rho_0}\Big{)}^{1 -
 \gamma_{\rho}}\;
\end{eqnarray}
where in the last step we assume $\gamma_{\rho} > 1$ and $\rho_{\rm
max}  \gg \rho$. For simplicity, factors that are of  order 1 such as $1 -
 \gamma_{\rho}$ are dropped from further analysis. Note that $\gamma_{\rho} >
1$ is a necessary condition for the integration to converge (which correspond to  $\gamma_{\rm N} > 1$). This is in
generally satisfied for the observed star-forming regions
\citep{2014Sci...344..183K,2015A&A...577L...6S}, and can be understood
theoretically \citep{2011ApJ...727L..20K}.

The mass of the region can be estimated using the shell approximation
\begin{equation}
{\rm d} M(\rho) \approx 4 \pi r^2\, \rho {\rm d} r\;,
\end{equation}
from which we derive (using $r \rightarrow 0 $ when $\rho$ is sufficiently
large) \footnote{  This quantity has been named as ``effective
radial density profile'', and has been discussed before. See
\citet{2011ApJ...727L..20K,2013ApJ...763...51F,2014Sci...344..183K}.}.
\begin{equation}\label{eq:r}
r(\rho) = \big (  \frac{P_{\rho_0}}{4 \pi} \big)^{\frac{1}{3}} \big{(}
\frac{\rho}{\rho_0}\big{)}^{- \frac{ \gamma_{\rho}}{3}} \;,
\end{equation}
which is
\begin{equation}\label{eq:rhor}
\rho =  \rho_0 \, \big (  \frac{P_{\rho_0}}{4 \pi }
\big)^{\frac{1}{\gamma_\rho}} r^{-\frac{3}{\gamma_\rho}} = \rho_0 \, \big (
\frac{P_{\rho_0}}{4 \pi } \big)^{\frac{1}{\gamma_\rho}} r^{-(1 +
\frac{2}{\gamma_{\rm N}})}\;.
\end{equation}
Observations have found that $\gamma_{\rm N}\approx 2$ which implies $\rho \sim
r^{-2}$ (see also
\citet{2011ApJ...727L..20K,2014Sci...344..183K,2014A&A...565A..24F,2013ApJ...763...51F}).

\subsection{multiscale gravitational energy}\label{sec:multi}
The gravitational binding energy of one single shell is
\begin{equation}
E_{\rm p}^{\rm shell} =    \frac{G M_{\rm enc}(\rho)\; {\rm d} M(\rho)
}{r(\rho)} \;,
\end{equation}
it can be simplified using Equations \ref{eq:d:mass}, \ref{eq:mass} and
\ref{eq:r},  where factors such $4 \pi$ are omitted for simplicity
\begin{equation}
E_{\rm p}^{\rm shell}  \approx \,  G\,  P_{\rho_0}^{\frac{5}{3}}\, \rho_0
\Big{(}\frac{\rho}{\rho_0}\Big{)}^{1 - \frac{5}{3} \gamma_{\rho}} {\rm d}\rho
\;,
\end{equation}
and
\begin{equation}
E_{\rm p}^{\rm shell}(r) \approx  G\; P_{\rho_0}^{\frac{1}{\gamma_\rho}}\,
\rho_0 \;r^{\frac{4 \gamma_{\rho} - 6}{\gamma_{\rho}}} |{\rm d} r|\;.
\end{equation}
Defining the wavenumber $k =  2 \pi /r$, following the convention used in
turbulence studies \citep{1995tlan.book.....F}, the \emph{gravitational energy
spectrum} of the system is,
\begin{equation}\label{eq:spectra}
E_{\rm p}(k) =  \Big{|}\frac{{\rm d}E_{\rm p}^{\rm shell} (k)}{{\rm d} k}\Big{|}
\approx   G\; P_{\rho_0}^{\frac{1}{\gamma_\rho}}\,
\rho_0
k^{\frac{-6 ( \gamma_{\rho} - 1)}{\gamma_{\rho}}}\;,
\end{equation}
where $E_{\rm p}(k)$ has a unit that is the same as the turbulence
power spectrum $E_{\rm turb}(k)$.
Using Eq. \ref{eq:ncoltorho}, we can express it as a function
of the scaling exponent $\gamma_{\rm N}$ of the surface density PDF
(where $P(N) \sim N_{\rm col}^{-\gamma_{\rm N}}$)
\begin{equation}\label{eq:final}
E_{\rm p}(k) \approx  G\; P_{\rho_0}^{ \frac{\gamma_{\rm N} + 2}{3 \gamma_{\rm
N}}}\,
\rho_0 \;
k^{-4(1 - \frac{1}{\gamma_{\rm N}})} \sim k^{ -4(1 - \frac{1}{\gamma_{\rm
N}})}\;.
\end{equation}

\begin{figure*}
\includegraphics[width=0.95 \textwidth]{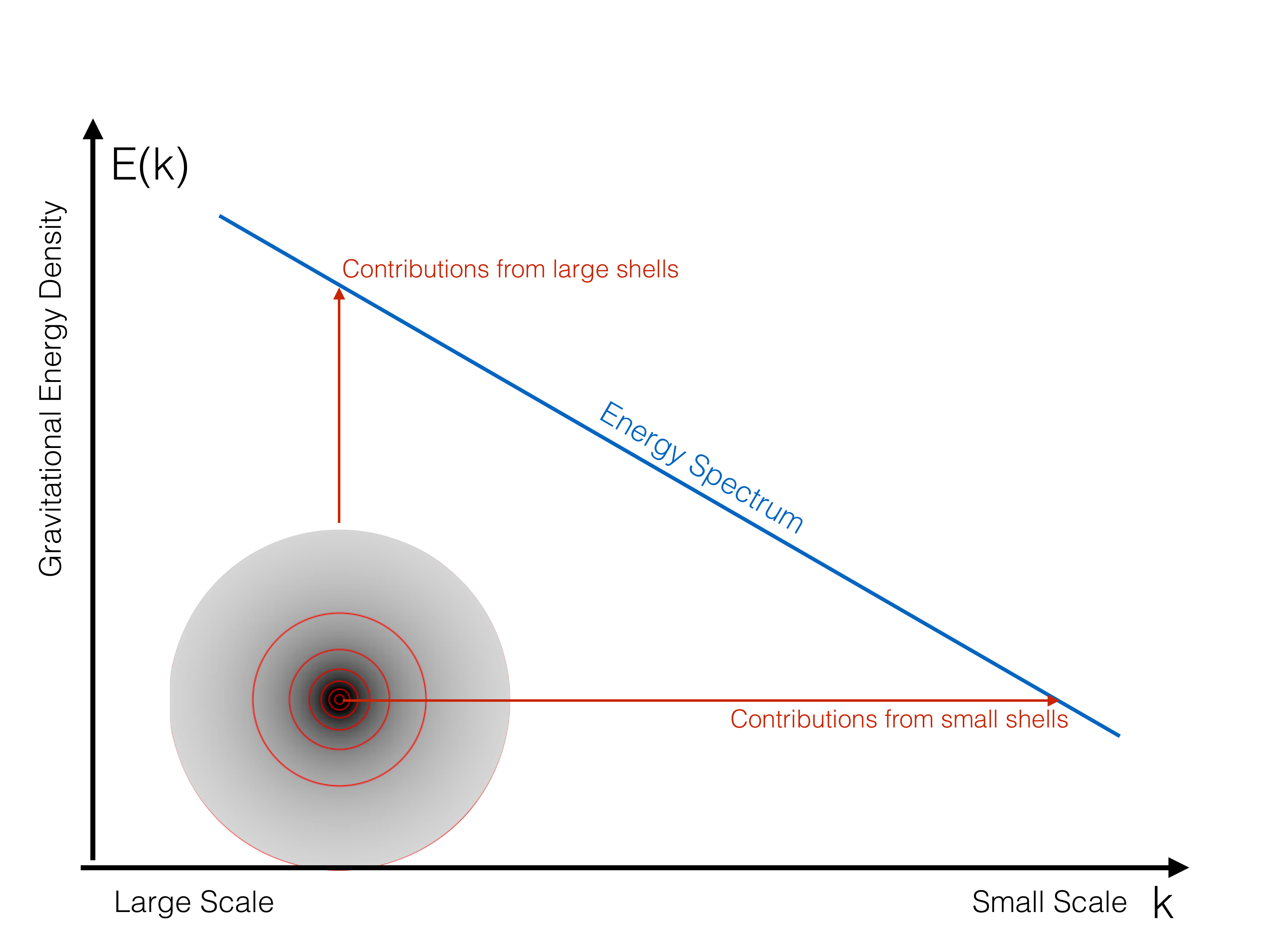}
\caption{A illustration of the concept of gravitational energy spectrum. 
{  The image on the bottom left is produced from an idealised cloud,
where we divide it into a set of nested shells. The boundaries of the shells
are marked by the red contours. The formula for making this density structure 
can be found in the clump model section of \citet{2016arXiv160305720L}. } The
gravitational energy spectrum is the distribution of the total gravitational
energy of the cloud at different wavenumbers $k$ where $k$ is related to the physical scale by $k = 2 \pi /l $.
A large $k$ represents the gravitational energy at a small scale, contributed
from gas that resides in the inner parts of the region. A small $k$ represent
the gravitational energy at a large scale, contributed by gas at the outer
envelopes.
See Sec. \ref{sec:multi} for details.\label{fig:ek:shell}}
\end{figure*}

Here we briefly explain the meaning of the formula for the gravitational energy
spectrum (Eq. \ref{eq:final}). The gravitational energy spectrum is the
distribution of the total gravitational energy of the cloud at different
wavenumbers $k$ where $k$ is related to the physical scale by $k = 2 \pi /l $.
A large $k$ represents the gravitational energy at small scales, contributed
by gas that resides in the inner parts of the region. A small $k$ represents the
gravitational energy at large scales, contributed by gas at the outer
envelopes.
 A steeper slope of $E_{\rm p}(k) $ means
gravitational energy is concentrated at larger scales, and a shallower slope
means gravitational energy is concentrated at smaller scales. This is
illustrated in Fig. \ref{fig:ek:shell}.

The energy spectrum satisfies energy
conservation
\begin{equation}
\int E_{\rm p}(k) {\rm d} k = \int E_{\rm p}(r) {\rm d} r = E_{\rm p}^{\rm
tot}\;,
\end{equation}
where $E_{\rm p}^{\rm tot}$ is the total gravitational binding energy of the
system.
Here we prefer to use the wavenumber $k$ over the size $l$ as the gravitational
energy spectrum should have an identical form to the turbulent kinetic energy
spectrum $E_{\rm turb}(k)$.

\subsection{Conversion between N-PDF, $\rho$-PDF and $\rho(r)$ and $E_{\rm
p}(k)$}\label{sec:formulas} The formalism described above enables us to convert
the observed N-PDF into $\rho$-PDF and finally into the density profile
$\rho(r)$ and derive the gravitational energy spectrum $E_{\rm p}(k)$. This
enables the observers to interpreted the observed surface density PDF with
 physically meaningful models. For this purpose, we collected all the useful
 formulas below. These formula are accurate in order-of-magnitude sense. Suppose
 we have a region of size $L$, and this region has a minimum surface density $N_0$. In this region, the high surface density part stays
inside envelopes of low surface densities (like the case of NGC1333 shown in
Fig. \ref{fig:model}),
 and its N-PDF can be written as
\begin{equation}
P(N_{\rm col}) = L^2 \Big( \frac{N_{\rm col}}{N_{0}} \Big)^{- \gamma_{\rm
N}}\;,
\end{equation}
where $N_0$ is a critical surface density.
According to the \emph{shell model}, its $\rho$-PDF is (from Eq. \ref{eq:prho})
\begin{equation}
P(\rho)  = L^3 \times \Big(\frac{\rho}{\rho_0}\Big)^{- \gamma_{\rho}}\;,
\end{equation}
where $\rho_0\approx N_0 / L $ and $\gamma_{\rho}$ is given by Eq.
\ref{eq:ncoltorho}. One can derive the radial density profile (from Eq.
\ref{eq:rhor}):
\begin{equation}
\rho(r)  =  \rho_0 \, 
L^{\frac{3}{\gamma_\rho}} r^{-(1 +
\frac{2}{\gamma_{\rm N}})}\;.
\end{equation}
The gravitational energy spectrum of the system is (from Eq. \ref{eq:final})
\begin{equation}
E_{\rm p}(k) \approx  G\; L^{ \frac{\gamma_{\rm N} + 2}{\gamma_{\rm
N}}}\,
\rho_0 \;
k^{-4(1 - \frac{1}{\gamma_{\rm N}})} \sim k^{-4(1 - \frac{1}{\gamma_{\rm
N}})}\;.
\end{equation}

\subsection{Uncertainty from the underlying geometry of the gas}
Star-forming regions are known to be irregular, having substructures. In our
simplified \emph{shell model}, the cloud is  approximated as a collection of
nested shells. This will introduce some inaccuracies in the estimate of
gravitational energy. Here we briefly discuss this accuracy issue.

We consider a thought experiment where we artificially split an object into $N$
completely separated identical sub-clumps and keep the density distribution
unchanged.
The gravitational energy of an object of mass $M$ and size $r_0$ is
\begin{equation}
E_0  =  \eta \frac{G M^2}{r_0}\;,
\end{equation}
where $\eta \sim 1$ and is dependent on the underlying geometry of the gas.
After the artificial fragmentation, this object is divided into  $N$ smaller
objects of equal mass of $M/N$ and equal radius $r_0 /N^{1/3} $. The total
gravitational energy of the artificially fragmented system is
\begin{equation}\label{eq:n}
E_1 =  \eta \frac{G (M/N)^2 \times N}{r / N^{1/3}} = E_0 N^{-2/3}\;.
\end{equation}
Thus this artificial fragmentation process will decrease the gravitational
binding energy by a factor of $N^{-2/3}$ where $N$ is the number of subregions.

The energy difference created by this artificial splitting (fragmentation)
process can be considered as a safe upper limit to the error of the estimation
in reality. This is because in our calculations, after the artificial fragmentation experiment,
the clumps are assumed to be gravitationally non-interacting. But in reality
they are gravitationally interacting. This will decrease the energy
difference between $E_0$ and $E_1$. Practically speaking,
in many cases, we are interested in the general slope of the gravitational
energy spectrum, as it tells us
how the gravitational energy evolves with scale. { In these cases,
 $N$ only influences the normalization of the gravitational energy 
spectrum, and does not change the slope.}

\begin{figure*}
\includegraphics[width = 0.44 \textwidth]{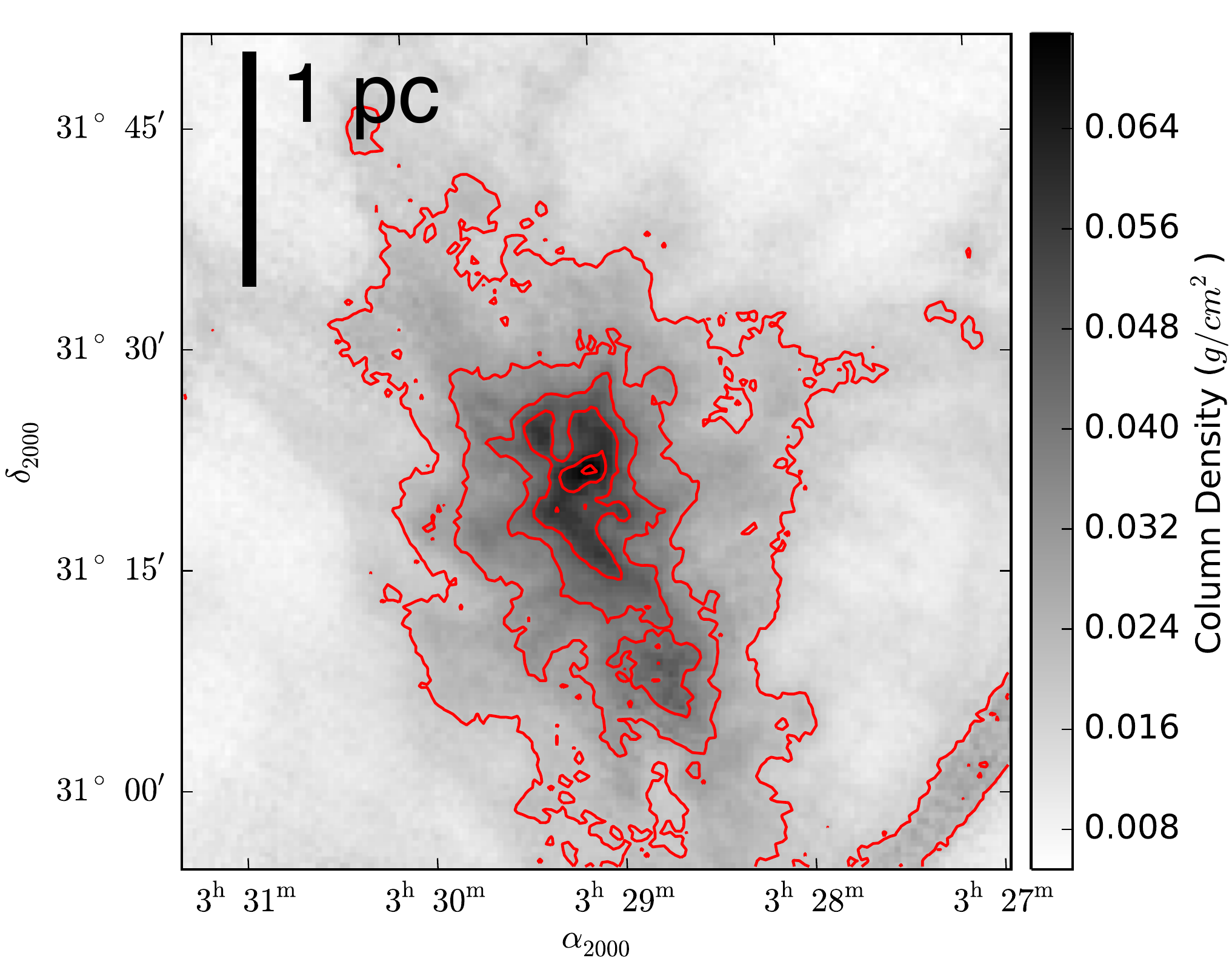}
\includegraphics[width = 0.52 \textwidth]{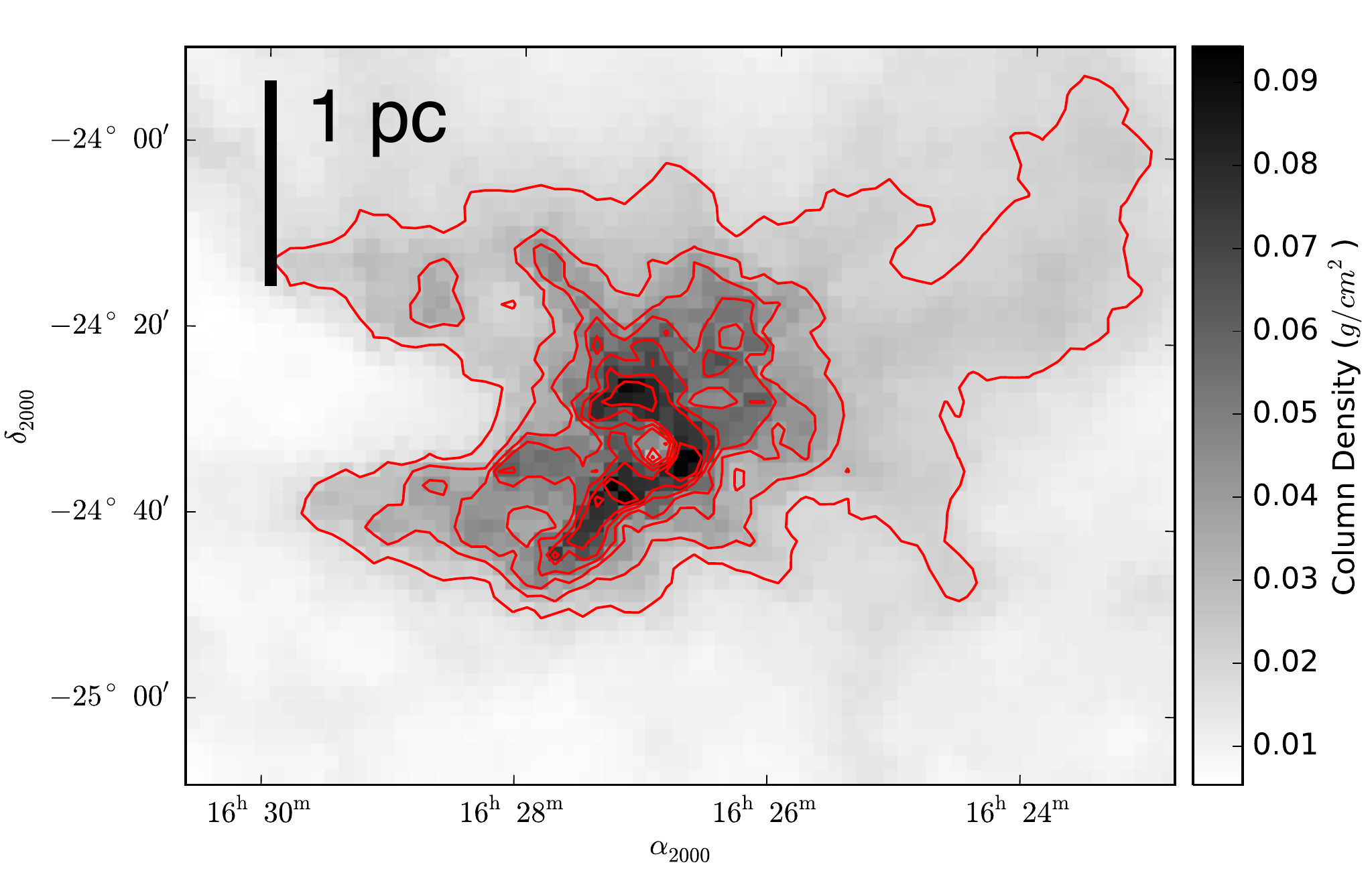}
\caption{Surface density map of two star-forming regions.  The NGC1333 region is
displayed on the left panel. The surface density map is produced from the
$^{13}$CO(1-0) map, using a conversion of $n_{\rm H_2} = 5 \times 10^{20}\rm\;
(K\; \rm km/s)^{-1}$. The Ophiuchus star-forming region is displayed on the
right panel.
The surface density is estimated from extinction, using $n_{\rm H_2} = 9\times
10^{20}\;A_{\rm v}^{-1}$ using COMPLETE \citep{2006AJ....131.2921R} data.
{  The complexities of these regions can be seem from the wedged
contours that traces the surface density distributions. High
density regions tend to reside in envelopes of lower densities. This feature is
captured in our model.}
\label{fig:obs}}
\end{figure*}

Typically (see Fig. \ref{fig:obs}), for compact star forming regions like
NGC1333 in the Perseus molecular cloud, the high density parts always stay inside nested
envelopes of lower densities.
Therefore
 $N\approx 1$ provides a fairly good description of the geometrical
structure of such regions. For the Ophiuchus molecular cloud, fragmentation
occurs, {  however, because the fragments are still close to each
other spatially, and are probably interacting with each other gravitationally,
we expect our model to be accurate. For regions like the Perseus, ideally, one could separate it into into subregions (e.g.
NGC1333, B1, B2 and IC348) and evaluate the surface density PDFs and
gravitational energy spectra for these regions. Alternative, by assume that
these regions are almost identical, Eq. \ref{eq:final} can still be used to derive the slope of the  gravitational energy spectrum. In this case, the normalization should
 be modified according to Eq. \ref{eq:n}.}
 Clouds like the Polaris molecular cloud are composed of many sub-regions. Here, one can still separate
the cloud into individual subregions and evaluate the gravitational energy
spectra of these regions, respectively. However, it is more convenient to use Eq. \ref{eq:final}
to derive the gravitational energy spectrum of the cloud as a whole. In this case, it is implicitly assumed that these regions have somewhat similar shapes.

{ 
The geometry of star-forming regions are not symmetric, and filamentary
structures has been seen on almost all scales}
 \footnote{There are different filaments.  See \citep{2011A&A...529L...6A} for
 filament network inside the cloud. A comprehensive list of
 filaments larger than the cloud scale has been collected in \citep{2016arXiv160400544L}. }.
This might also be a contributing factor to the inaccuracy of the model. However, a detailed calculation suggests that this effect is not significant. The simplest way to
quantify this is to consider the impact of aspect ratio on the estimated
gravitational energy.
An aspect ratio of $\sim 10$ only changes the gravitational binding energy by a
factor of $\sim 1.5$ (see Appendix \ref{sec:bertoldi} for details).
Thus our Equation \ref{eq:spectra} should be a good approximation
of the energy spectrum in spite of all the above-mentioned compilations. \footnote{  If the filament have a uniform density and is infinitely
long, one need to use cylindrical model instead of the shell model. However,
this special case is too artificial and has never been seen observationally. In
most cases, the filaments are already fragmented, and we expect our model to be
applicable to the fragmented filaments. } 
{ 
When the regions are too
complicated to be approximated with the shell model, one can also
construct the gravitational energy spectrum numerically
(see an accompanying paper,
\citep{2016arXiv160305417L}). In fact, using data from
the Orion molecular cloud,  \citet{2016arXiv160305417L} demonstrated that the
shell model appears to be a good approximation.
}

\section{Interplay between turbulence and gravity }\label{sec:obs}
\subsection{Observational results}
\begin{figure*}
\includegraphics[width = 0.95 \textwidth]{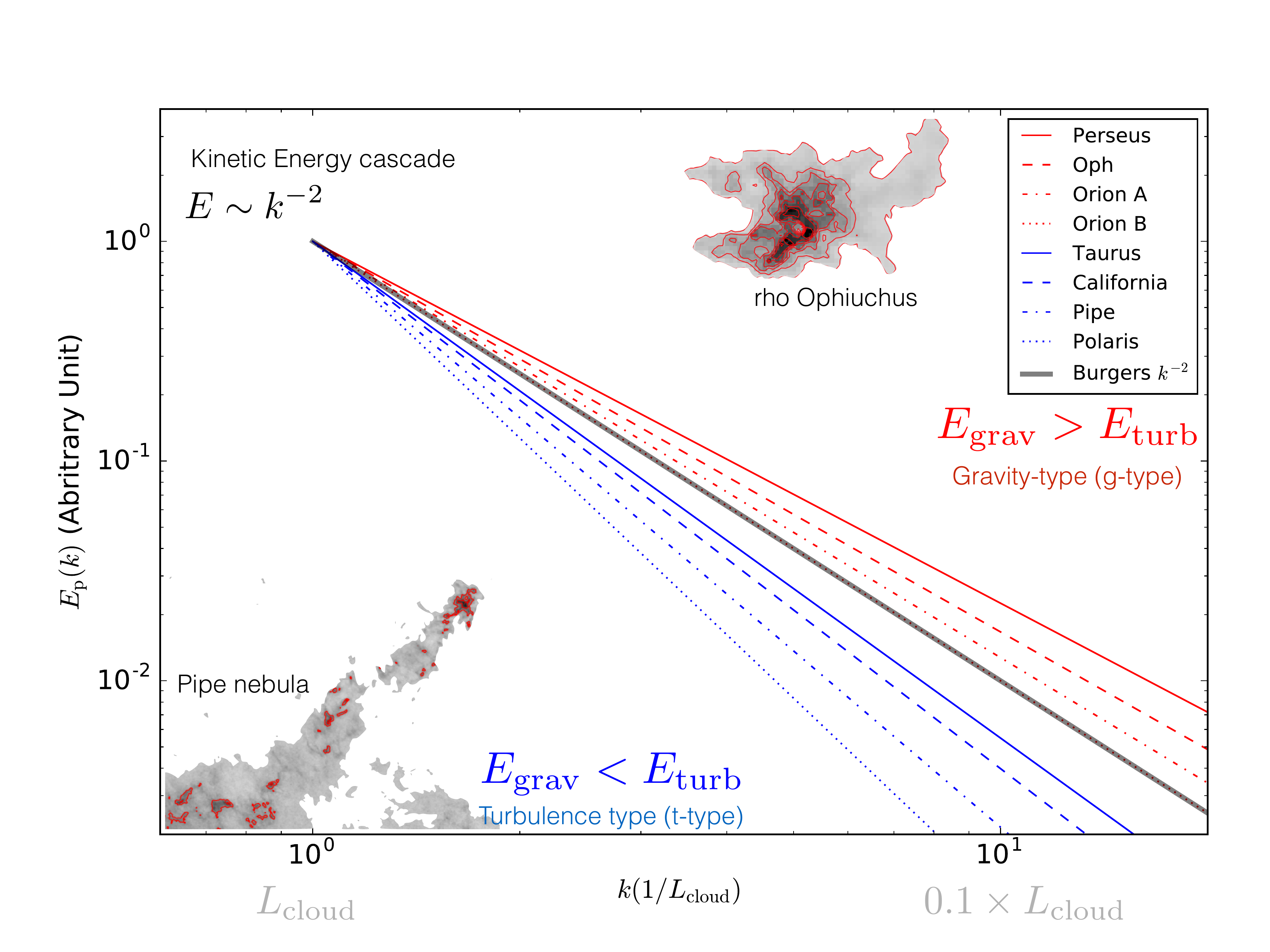}
\caption{The derived gravitational energy spectra of several molecular clouds.
The energy spectra are derived from Table 1 of
\citet{2015A&A...576L...1L} (see also our Table \ref{tbl:properties}), using
Eq. \ref{eq:final}. Gravitational energy at smaller scales (large $k$) is
contributed by the inner shells, and gravitational energy at larger scales
(small $k$) is contributed by the outer shells. {  The $x$-axis is
normalized to the size of the cloud $L_{\rm cloud}$ (which is around a few
parsec in size).
The derived gravitational energy spectra are valid from the cloud scale to the map resolution (which is better
than $ 0.1\, L_{\rm cloud}$ in our case).
The
normalization of the $y$-axis is arbitrary. }
Critical energy spectrum of Burgers turbulence $E(k)\sim k^{-2}$ is indicated by
the thick black line. The gravitational energy spectrum of the Orion B cloud have $E(k)\sim k^{-2}$ and coincides with that of
 Burgers turbulence.
 Above the critical energy spectrum, the
gravitational energy exceeds the inferred turbulence energy and gravitational
collapse dominates.
Below it, gravity can  be  balanced by turbulence. {  The embedded maps of
the Pipe nebula (represent our \emph{t-type} cloud ) is produced using the
extinction data of
\citet{2009MNRAS.395.1640R} and the $\rho$-Ophiuchus data (representing our
\emph{g-type} cloud) is from
\citet{2006AJ....131.2921R}. For clarity, we have chosen a normalization such
that the lines overlap at $L_{\rm cloud}$.}
\label{fig:result}}
\end{figure*}

It has been demonstrated observationally that beyond a threshold column density,
molecular clouds exhibit power-law PDFs \citep{2009A&A...508L..35K,2011A&A...530A..64K}. 
\citet{2015A&A...576L...1L} studied the surface density distribution of 8
nearby molecular clouds, and argue that the threshold surface density extends
down to $A_{\rm k}
\gtrsim 0.3$, which is lower than the value found in
\citet{2009A&A...508L..35K,2011A&A...530A..64K}. Different regions have
different scaling indexes. We use these scaling indexes to determine the gravitational energy spectra of the clouds.

\subsection{Two regimes of cloud evolution}

It is found that turbulence in molecular clouds is supersonic. The power
spectrum of the turbulence is believed to be close to that of Burgers
turbulence, which is $E_{\rm k} \sim k^{-2}$ \citep{2013MNRAS.436.1245F}. This is the
maximum amount of energy one would expect to be transferred to smaller scales
from turbulence cascade.
On the largest scale, molecular clouds are close to be gravitational bound \citep{2010ApJ...723..492R, 2009ApJ...699.1092H}, and the amount of turbulence
energy is comparable to the gravitational binding energy.
{  We note that there is still a large uncertainty (0.5 -- 5)
concerning the estimated virial parameters in the literature
 \citep{2007ApJ...661..830R,2015ApJ...809..154H}.
 \citet{2015A&A...578A..97L} have demonstrated that by carefully choosing the
boundaries of the regions, the cloud is much more gravitationally bound what is
suggested by a simple virial analysis. It remains to be investigated
 (following \citet{2015A&A...578A..97L}) if these uncertainties arise because of
 the chosen boundaries of the cloud, or the cloud are intrinsically unbound.}
 Many of the clumps in the clouds are in virial equilibrium
\citep{2012A&A...544A.146W}. 
The importance of gravity in massive star-forming
clumps has also been observationally demonstrated
using the $\alpha_{vir}-\alpha_{G}$ formalism \citep{2015arXiv151103670T}.

{ 
At high surface density, it has been noticed that the surface density PDF of
molecular clouds can be described by power-laws
\citep{2013ApJ...766L..17S,2012A&A...540L..11S,2009A&A...508L..35K}.  With an
almost-homogeneous sample of 8 molecular clouds, \citet{2015A&A...576L...1L}
found that the power-law surface density PDFs starts at at $A_{\rm v} \gtrsim
0.3$. The derived scaling exponents are scatters around -2.} Using Equation
\ref{eq:final}, this corresponds to a gravitational energy spectrum of $E_{\rm p}\sim k^{-2}$, which coincides exactly
with the energy spectrum of Burgers turbulence. In this case, if turbulence
can cascade effectively, the cloud should be in a critical state where the
turbulence and gravitational energy are comparable on  multiple  scales
(from a few parsec to subparsec).
A steeper gravitational
energy spectrum means that there is less gravitational energy than turbulent
energy; and a shallower
spectrum implies the dominance of gravitational energy.

In Fig. \ref{fig:result} we present the derived gravitational energy spectra using
the results from Table 1 of \citet{2015A&A...576L...1L}.
The results are valid from $A_{\rm k} \approx
0.3$ to $A_{\rm k} \lesssim 10$. The
gravitational energy spectra are evaluated using Eq. \ref{eq:final}.

\begin{table}
\begin{tabular}{ l|l l l }
\hline
Region Name & $\gamma_{\rm N}$ & $\gamma_{\rho}$ & $\gamma_{\rm E_{\rm p}}$ \\
  \hline
Perseus  & 1.7 & 1.38 & 1.65\\
Oph  & 1.8  & 1.42 & 1.78\\

Orion A &  1.9  & 1.46 & 1.89 \\
 Orion B  & 2.0  & 1.50 & 2.0\\
 \hdashline
 Taurus  & 2.3 & 1.60 & 2.26\\
 California & 2.5  & 1.67 & 2.4\\
 Pipe  & 3.0 & 1.80 & 2.66\\
Polaris &  3.9  & 1.98 & 2.97\\
\hline
\end{tabular}
\caption{ Properties of different regions. $\gamma_{\rm N}$ is the scaling exponent of the
observed surface density PDF  ($P(N_{\rm col}) \sim N_{\rm col}^{-\gamma_{\rm
N}}$ where $N_{\rm col}$ is the observed surface density).
$\gamma_{\rho}$ is the scaling exponent of the theoretically-derived volume
density PDF  ($P(\rho) \sim \rho^{-\gamma_{\rm \rho}}$). $\gamma_{\rm E_{\rm
p}}$ is the scaling exponent of the gravitational energy spectra ($E_{\rm p}(k)\sim k^{-\gamma_{\rm E_{\rm p}}}$).
$E_{\rm p}(k) \sim k^{-2}$ is
the theoretical boundary between gravity-dominated (g-type) clouds and
turbulence-dominated (t-type) clouds.
Clouds above this line are gravity-dominated and clouds below
this line are turbulence-dominated.
\label{tbl:properties} }
\end{table}

Molecular clouds like Perseus and Ophiuchus have shallow surface density PDFs
(where the scaling exponent is larger than -2). This corresponds to
gravitational energy spectra that are relatively flat compared to the case of
Burgers turbulence. If the clouds are gravitational bound on the large scale
$l_0$, it is inevitable that at any smaller scale ($l < l_0$, typically parsec
scale), gravitational energy will dominate turbulent energy.
Thus these regions are either undergoing gravitational
collapse
\citep[e.g.][]{2004ApJ...616..288B,2011MNRAS.411...65B,1953ApJ...118..513H,1993ApJ...419L..29E},
or are supported by e.g. magnetic fields.
Indeed, all these clouds are actively forming stars. The Orion
B molecular cloud  has a steeper gravitational energy spectrum ($\gamma_{\rm
E_{\rm p}} = 2$) as compared to Orion A ($\gamma_{\rm
E_{\rm p}} = 1.89$), which might explain why star formation in Orion B is $\sim
3$ times less efficient as compared to Orion A \citep{2016AJ....151....5M}.
It is interacting with winds from a collection of massive stars and is
probably close to disruption \citep{2008hsf1.book..459B}. Generally speaking,
for the category of clouds with $\gamma_{\rm E}<2$, gravity tends to dominate
over turbulence at smaller scales, we name this type as  gravity-dominated type (\textit{g-type}).

At the lower part of Fig. \ref{fig:result}. {  Our size scale is normalized with respect the sizes of the
individual regions (called $L_{\rm cloud}$ in Fig. \ref{fig:result}), which are
typically a few parsec in size, down to the map resolution, which is $\sim 0.1
\rm pc$. } The clouds have gravitational
energy spectra that are steeper than $k^{-2}$. In these cases, if turbulence can
cascade effectively onto the smaller scales, it can support the
cloud against gravitational collapse and dominates over gravitational forces.
However, as the energy cascade of supersonic turbulence under the influence of gravity is still not well
understood yet, we are refrained from drawing a firm conclusion. The clouds in
this regime can for example be supported magnetically.
We call this type of cloud turbulence-dominated type (\textit{t-type}).
 It should also be noted
that a cloud might belong to the marginal type, e.g. the Orion B
molecular cloud $E_{\rm p} \sim k^{-2}$, and is at the boundary between
turbulence-dominated and gravity-dominated types.

This distinction draws further supports from observational
studies of their star-formation activities. For \textit{t-type} clouds, apart from the
Taurus molecular cloud which has a power law index of the gravitational energy
spectrum of $\sim -2.26$, none of these clouds are considered as active in star formation.
{  (e.g. the \emph{t-type} California cloud has 10 times lower star
formation efficiency than Orion \citep{2009ApJ...703...52L}, and furthermore,
the Pipe nebula and Polaris are almost devoid of star formation).}
 The Taurus molecular cloud
is an interesting marginal case. It is considered as star-forming. The star
formation rate in Taurus is similar to that of the Ophiuchus molecular cloud
\citep{2012ApJ...745..190L}. However the star formation is extremely
distributed as compared to the clustered fashion in Ophiuchus.
Perhaps the lack of small-scale gravitational energy is directly related to the lack of clustered star formation in this cloud.
The Taurus molecular cloud is composed of two parallel filaments, where gravity
can probably trigger collapse in a non-uniform fashion
\citep{2004ApJ...616..288B,2016arXiv160305720L}.
 It
has also been suggested that Taurus molecular cloud is supported by feedback
\citep{2015ApJS..219...20L} and magnetic fields \citep{2008ApJ...680..420H},
consistent with our \textit{t-type} classification.

Note also that in the \textit{t-type} clouds, turbulence can provide
support against gravity. However, this does not necessarily mean that the
regions are unbound.
Since we only consider the bulk amount of gravitational energy,  even if one has
demonstrated that $E_{\rm turb} > E_{\rm p}$, some sub-regions can still be
gravitational bound. But to compensate this, other regions need to be unbound
in order  to accommodate this excess of kinetic energy.

{ 
Thus, our gravitational energy spectrum allows one to related the observed
surface density PDF to the important of gravity in the clouds. Since on small
scales, star formation is dominated by gravity, we expect a direct connection
between the surface density PDF and star formation activity. In fact, this has
been demonstrated observationally in 
\citet{2014Sci...344..183K}, where the slopes of the
density PDF correlate with the star formation efficiency.}

Overall, the molecular clouds as studied in \citet{2015A&A...576L...1L} and this
work have energy spectra that scatter around the critical value of $\gamma_{\rm
E}=2$ for which turbulence and gravity can balance each other. This
suggests that in general, there exists multiscale equipartition between gravity
and turbulence.
However, the variations of gravitational energy spectrum are still significant:
assuming that all these clouds are roughly gravitational bound on the cloud scale ($l_0$), their gravitational energy per mass can differ by more
than one order or magnitude on smaller scales ($\approx 0.1\; l_0$). This
significant difference indicates that molecular clouds can belong to two
separate categories (or two states \citep{2012ApJ...750...13C}): in the
\textit{g-type} clouds, gravity can dominate over turbulence and in the other;
and in the \textit{t-type} clouds, turbulence could provide support against
gravity.

\section{Conclusions \& Discussions}\label{sec:conclu}
In this work, by approximating the observed star-forming regions as collections
of spherically symmetric nested shells where gas of high densities resides in
envelops of lower densities, we derive an analytical formula for the
gravitational energy spectrum.
If, above a minimum surface density $N_{\rm min}$, a cloud has a density PDF of
the form $N= N_0\; N_{\rm col}^{- \gamma_{\rm N}}$, it can be approximated
as a set of nested shells that are described by
:
\begin{equation}\label{eq:final:f}
\rho(r)  =  \rho_0 \, \big (
\frac{L^3}{4 \pi } \big)^{\frac{1}{\gamma_\rho}} r^{-(1 +
\frac{2}{\gamma_{\rm N}})}\sim r^{-(1 +
\frac{2}{\gamma_{\rm N}})}\;,
\end{equation}
and the
gravitational energy spectrum is
\begin{equation}\nonumber
E_{\rm p}(k) \approx  G\; L^{ \frac{\gamma_{\rm N} + 2}{\gamma_{\rm
N}}}\,
\rho_0 \;
k^{-4(1 - \frac{1}{\gamma_{\rm N}})} \sim k^{-4(1 - \frac{1}{\gamma_{\rm
N}})}\;,
\end{equation}

where $\rho_0 \approx N_0 / L$, $L$ is the size of the
region. The wavenumber $k$ is $k = 2 \pi
/ l$ where $l$ is the length scale. For a typical molecular cloud with
$\gamma_{\rm N} \approx 2$, it satisfies $\rho\sim r^{-2}$, which implies a
gravitational energy spectrum of $E_{\rm p}(k)\sim k^{-2}$.

Eq. \ref{eq:final:f} enables us to evaluate the distribution of gravitational
energy over multitude scales. This can be compared with the expected
kinetic energy distribution from e.g. turbulence cascade.
Cascade of Burgers turbulence gives $E_{\rm k}\sim k^{-2}$.
Since molecular clouds are
found to have $E_{\rm p}\sim k^{-2}$, this implies
a equipartition between turbulence and gravitational energy across different
 scales.

By investigating the gravitational energy spectra of individual molecular clouds
in details, we find that molecular clouds can be broadly divided into two
categories. The \textit{g-type} includes the clouds
shallow surface density PDFs ($\gamma_{\rm N} < 2$, including e.g. the Persues
and the Orion A molecular cloud).
Inside these clouds, on smaller scales, the gravitational energy exceeds by much
the turbulence energy from cascade. As a result, it is difficult for turbulence to support these clouds. Either they are
experiencing gravitational collapses
\citep{2011MNRAS.411...65B,1953ApJ...118..513H,1993ApJ...419L..29E,2004ApJ...616..288B},
or they are supported by other physics, such as  magnetic fields
\citep{2014prpl.conf..149T,2014prpl.conf..101L}.
The \textit{t-type} includes clouds with steep slopes of the surface
density PDFs ($\gamma_{\rm N} > 2$, for example, the Pipe nebula and the California
molecular cloud).
For them, the bulk turbulence energy exceeds the gravitational energy at small
scales, and the sub-regions in these clouds can be supported by turbulence.
This theoretical distinction is supported by observations, in that the first
type of clouds (the g-type) are forming stars actively, and the second type 
(the t-type) are relatively quiescent.

The fact that gravity takes over turbulence on every given physical scale for
clouds like Orion A is interesting and  deserves further
investigations e.g. \citet{2013ApJ...773...48B}.
There are different models of cloud evolution.
Because of the
large amount of gravitational energy distributed across various physical scales,
 many of the  turbulent motions in molecular cloud can and should
 be explained as a result of gravity
 \citep{2009ApJ...699.1092H,2015arXiv151105602I,2011MNRAS.411...65B,2015arXiv151103670T}.

Our results provide a refined, multiscale picture of gravity in cloud
evolution. The analytical formulas in Sec. \ref{sec:formulas} are helpful for
interpreting observations that constrain column density PDFs.
Equation \ref{eq:final} can be used to convert observed column density
distributions into the multiscale gravitational energy spectrum.
It leads to a
critical theoretical criterion to separate molecular clouds into  two distinct
types (turbulence-dominated t-type and gravity-dominated g-type),
and one need to understand the development of gravitational instability in these
different regimes e.g. \citep{1995A&A...303..204V,1987A&A...172..293B}.  A unified theory of star formation should take this
distinction into account.

\section*{Acknowledgements}
Guang-Xing Li thanks Alexei Kritsuk, Martin Krause and Philipp Girichidis for
discussions.
Guang-Xing Li is supported by the Deutsche Forschungsgemeinschaft (DFG) priority
program 1573 ISM- SPP. We thank the referee for a careful reading of the paper
and the helpful comments. 
The paper also benefits from comments  Jouni Kainulainen,
and Christoph Federrath.

\appendix

\section{Influence of the aspect ratio}\label{sec:bertoldi}
The formula for gravitational energy of a 3D ellipsoid of sizes $(R, R, Z)$ and
mass $M$ has been derived by \citet{1992ApJ...395..140B}
\begin{equation}
E_{\rm p} = \frac{3}{5}\frac{G M}{R^2}\frac{{\rm arcsin}(e)}{e}
\end{equation}
where $e$ is the eccentricity ($e = \sqrt{1 - 1/y^2}$, $y = R/Z$ is the aspect
ratio). An change in the aspect ratio by a factor of 10 only changes the
gravitational energy by a factor of $\approx 1.4$. The dependence of
gravitational energy on aspect ratio is extremely weak.

\bibliography{paper}
\end{document}